\documentclass[times,authoryear,twocolumn,final]{elsarticle}

\usepackage{medima}
\usepackage{framed,multirow}
\usepackage{amssymb}
\usepackage{latexsym}
\usepackage{gensymb}
\usepackage{amsmath}
\usepackage{epsfig}
\usepackage{url}
\usepackage{xcolor}
\usepackage{lmodern}
\usepackage{hyperref}

\usepackage{caption,subcaption}

\def\ddict{\mathcal{D}}
\def\embed{\mathcal{E}}
\def\tsne{\mbox{t-SNE}}
\def\dIR{\ddict_\text{IR}}
\def\dTSE{\ddict_\text{TSE}}
\def\dJ{\ddict_\text{J}}
\def\dS{\ddict_\text{S}}
\def\eIR{\embed_\text{IR}}
\def\eJ{\embed_\text{J}}

\DeclareMathOperator*{\argmax}{argmax}

\graphicspath{{./Figures/}}

\journal{Medical Image Analysis}

\begin{document}

\verso{Kirsten Koolstra \textit{et~al.}}

\begin{frontmatter}

\title{Hierarchical stochastic neighbor embedding as a tool for visualizing the encoding capability of magnetic resonance fingerprinting dictionaries}%

\author[1]{Kirsten \snm{Koolstra}}
\author[1,2]{Peter \snm{B\"{o}rnert}}
\author[3,4]{Boudewijn P.F. \snm{Lelieveldt}}
\author[1]{Andrew \snm{Webb}}
\author[3]{Oleh \snm{Dzyubachyk}\corref{cor1}}

\address[1]{C.J. Gorter Center for High Field MRI, Department of Radiology, Leiden University Medical Center, Albinusdreef 2, 2333 ZA, Leiden, The Netherlands}
\address[2]{Philips Research Hamburg, R\"{o}ntgenstrasse 24, 22335, Hamburg, Germany}
\address[3]{Division of Image Processing, Department of Radiology, Leiden University Medical Center, Albinusdreef 2, 2333 ZA, Leiden, The Netherlands}
\address[4]{Intelligent Systems Department, Delft University of Technology, Mekelweg 4, 2628 CD, Delft, The Netherlands}
\cortext[cor1]{Corresponding author.
  Tel.: +31\,715261911;
  \ead{o.dzyubachyk@lumc.nl}}

\begin{abstract}
In Magnetic Resonance Fingerprinting (MRF) the quality of the estimated parameter maps depends on the encoding capability of the variable flip angle train. In this work we show how the dimensionality reduction technique Hierarchical Stochastic Neighbor Embedding (HSNE) can be used to obtain insight into the encoding capability of different MRF sequences. Embedding high-dimensional MRF dictionaries into a lower-dimensional space and visualizing them with colors, being a surrogate for location in low-dimensional space, provides a comprehensive overview of particular dictionaries and, in addition, enables comparison of different sequences. Dictionaries for various sequences and sequence lengths were compared to each other, and the effect of transmit field variations on the encoding capability was assessed. Clear differences in encoding capability were observed between different sequences, and HSNE results accurately reflect those obtained from an MRF matching simulation.

\end{abstract}


\end{frontmatter}


\section{Introduction}

Magnetic Resonance Fingerprinting (MRF) is a rapid MRI technique that is used to estimate tissue relaxation times ($T_1$, $T_2$) and other MR-related parameters such as proton density ($M_0$) \citep{bib:Ma2013}. These parameters often reflect pathology such as inflammation and neurodegeneration. Unlike many other quantitative imaging techniques, MRF simultaneously encodes $T_1$ and $T_2$, such that the corresponding parameter maps can be obtained in an efficient manner. The simultaneous encoding is established through a variable flip angle pattern in the data acquisition process, which, if designed well, creates a characteristic signal evolution for each tissue in the human body. The $T_1$ and $T_2$ values for each voxel can be found by matching the measured signal curve to a pre-calculated dictionary containing the simulated signal evolutions as a function of the applied flip angle sequence for all possible ($T_1$,\,$T_2$) combinations.

The quality of the resulting parameter maps substantially depends on the underlying MRF flip angle sequence. Recent works have shown that flip angle pattern optimization can either improve the accuracy of parameter quantification or reduce the scan time that is needed to achieve the same accuracy \citep{bib:Sommer2017,bib:Zhao2018}. It is also known that increasing the length of the MRF sequence improves the accuracy of the parameter maps, in particular $T_2$ \citep{bib:Cline2017,bib:Sommer2017}. Therefore, determining the optimal sequence or flip angle train is very important.

However, the process of optimizing a sequence is not straightforward due to the large solution space and the lack of well-established measures of encoding quality. Moreover, the optimal sequence may actually be different for each application, and therefore the application of interest and its constraints should ideally be taken into account. \citet{bib:Sommer2017} have shown how a Monte-Carlo type approach can be used to predict the encoding capability of different MRF sequences. The measures of encoding are based on the inner product between neighboring dictionary elements, and the distinction is made between local and global measures of encoding. Later, \citet{bib:Cohen2017} and \citet{bib:Zhao2018} formulated the sequence optimization problem as an inverse problem, allowing one to actually calculate the optimized sequence under certain constraints, using a dot product matrix as the encoding measure. Although these techniques show promising results, it is not clear yet how the encoding capability changes within a dictionary when a single number is assigned to the (global or local) encoding capability of the dictionary.

In this work we present an alternative approach to judge the encoding capability of MRF sequences that provides insight into local capabilities as well as global capabilities of encoding. We analyze the encoding capability of an MRF sequence by looking at its corresponding MRF dictionary, describing the relevant signal evolutions for the application of interest, as was also done in \cite{bib:Sommer2017}. We use the dimensionality reduction technique Hierarchical Stochastic Neighbor Embedding (HSNE) \citep{bib:Pezzotti2016}, which is a scalable and robust implementation of the t-Distributed Stochastic Neighbour Embedding (t-SNE) \citep{bib:vanDerMaaten2008}, to transform the high-dimensional MRF dictionary into a low-dimensional space. The choice of HSNE is motivated by its capacity to pick up small differences in signals while preserving the manifold structure, which makes it particularly useful for analyzing data with nonlinear structure such as MRF dictionaries. This allows us to visualize the entire MRF dictionary as a colormap, based on which the local and global encoding capability can easily be examined. The color values in these maps are a surrogate for location in the low-dimensional space. Moreover, this method provides a framework for comparing different MRF dictionaries and hence sequences. Several targeted experiments were carried out to demonstrate how our visual representation of MR and MRF dictionaries correspond to simulation experiments.

In Section~\ref{ssec:results-classical} we show the principles of the technique by analyzing dictionaries generated from classical quantitative sequences such as inversion recovery (IR) and turbo spin echo (TSE). Next, in Section~\ref{ssec:results-MRF}, we show that this technique is sensitive enough to pick up small differences in the encoding of MRF dictionaries by comparing three different MRF sequences. In Section~\ref{ssec:results-length} we visualize how the encoding capability changes with the length of the MRF sequence, and, as a final example, in Section~\ref{ssec:results-B1} we also demonstrate how the effect of variations in the intensity of the MR transmit field ($B_1^+$) on the encoding quality can be analyzed. The HSNE results are compared to simulation results for validation.



\section{Methods}
In MRF, $T_1$, $T_2$ and $M_0$ values are calculated by matching the measured signal evolutions to a calculated dictionary using the inner product as a quality measure. This is different from traditional parameter quantification, for which there is a closed form signal equation that describes the signal intensity of the images. In the latter case, the measured signal evolutions can be fitted to the signal equation using least squares minimization methods, resulting in $T_1$, $T_2$, $T_2^*$ and $M_0$ estimations. Traditional parameter quantification, however, can also be approached as a dictionary matching problem if the closed-form signal equation is used as a model for the dictionary construction. Here, we create such dictionaries for traditional quantification methods, such that the encoding capabilities of classical sequences can be studied in a similar way as for the MRF sequences.

\subsection{Classical dictionaries}
\label{ssec:classical}
Two different classical sequences were used to generate three classical dictionaries: the TSE sequence used for $T_2$ mapping (dictionary $\dTSE$), and the IR gradient echo sequence used for $T_1$ mapping (dictionary $\dIR$). The IR sequence was also studied with a shorter maximal inversion time to reduce the degree of $T_1$ encoding (dictionary $\dIR^{short}$). Dictionaries for these sequences were generated from the closed form signal ($S$) expression for the TSE sequence,
\begin{equation}
S(\text{TE}) = M_0\left(1-e^{-\frac{\text{TR}}{{T}_1}}\right)e^{-\frac{\text{TE}}{{T}_2}},
\end{equation}
and for the IR sequence,
\begin{equation}
S(\text{TI}) = M_0\left(1-2e^{-\frac{\text{TI}}{{T}_1}}+e^{-\frac{\text{TR}}{{T}_1}} \right)e^{-\frac{\text{TE}}{{T}_2}}.
\end{equation}
$\dTSE$ was calculated with TR=1.5~s. For $\dIR$, TR=2.5~s and TE=2~ms was used and the maximal inversion time was set to TI{$_\text{max}$=3~s. $\dIR^{short}$ was generated with the same TE and TR, but with a shorter maximal inversion time: TI$_{max}$=200~ms. Signal curves were discretized into 1000 time points to match the number of flip angles used for calculation of the MRF dictionaries. For simplicity, we set $M_0=1$ in these calculations. $T_1$ values ranged from 20 to 5000~ms in steps of 30~ms, and $T_2$ values ranged from 10 to 1000~ms in steps of 10~ms.

\subsection{MRF dictionaries}
\label{ssec:MRF_dict}
Two different MRF flip angle sequences were used to generate three MRF dictionaries. Both sequences consist of 1000 flip angles using a constant TR=15~ms. The sequence shown in Figure~\ref{fig:figure1} contains a smoothly varying flip angle pattern introduced by \citet{bib:Jiang2015} (dictionary $\dJ$) and is preceded by a 180$\degree$ inversion pulse. The same sequence was also analyzed without the inversion pulse (dictionary $\dJ^{-}$) to reduce the $T_1$ encoding ability. The third sequence constructed by \citet{bib:Sommer2017} (dictionary $\dS$) has a more jagged random pattern; see also Figure~\ref{fig:figure1}. The three MRF dictionaries were created by Bloch simulations using the extended phase graph formalism to model a fast imaging with steady state precession (FISP) sequence (unbalanced) \citep{bib:Scheffler1999}. The same $T_1$ and $T_2$ ranges were used as for the classical dictionaries. For Jiang's pattern (with inversion pulse) a dictionary $\dJ^{B}$ was also generated, taking into account $B_1^+$ variations ranging from 0.4 to 1.3 times the nominal values to mimic the impact of transmit RF inhomogeneity. All dictionary calculations only included ($T_1$,\,$T_2$) combinations for which $T_1$ is larger than $T_2$.

\begin{figure}[!t]
\centering
\includegraphics[width=\columnwidth]{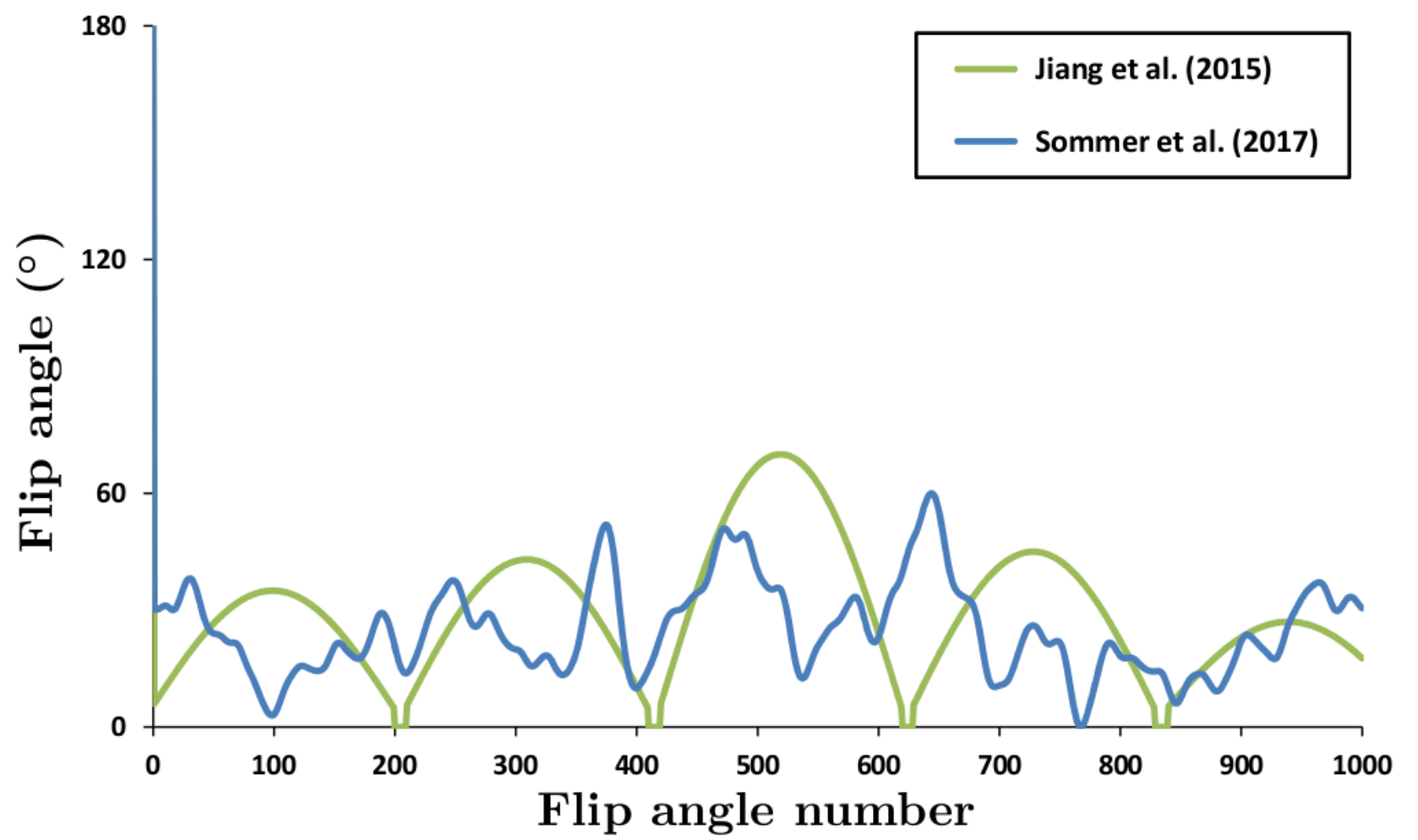}
\caption{\small MRF flip angle patterns used: smoothly varying pattern designed by \citet{bib:Jiang2015} (green line) and randomly varying pattern designed by \citet{bib:Sommer2017} (blue line). Both patterns start with an inversion pulse seen at flip angle number 0.}
\label{fig:figure1}
\end{figure}

\subsection{Dimensionality reduction}
Each dictionary entry was reduced from 1000 to either 2 elements (for the classical sequences) or 3 elements (for the MRF sequences) with \tsne, which projects higher-dimensional data onto a lower-dimensional manifold while preserving similarity (pairwise distances) between data points. We used HSNE as the particular implementation of \tsne\ as it has been shown to be much more capable of reconstructing the underlying low-dimensional manifolds \citep{bib:Pezzotti2016}. Embeddings were initialized with random seed placement. Drilling into the next (lower) level of the hierarchy was performed after the convergence ($10^5$ iterations) of the current level, using all the calculated landmarks. Landmarks that were added on the lower level were initialized by interpolating the locations of their ``parents''. Early exaggeration (200 iterations; factor 1.5) was used for producing the embedding only on the top level of the hierarchy. The embedding on the bottom (data) level was selected as the final result; see examples in Figures~\ref{fig:classical}--\ref{fig:lengths}. No additional data standardization (e.g. Z-scoring) was performed.

\begin{figure}[!tb]
\includegraphics[width=\columnwidth]{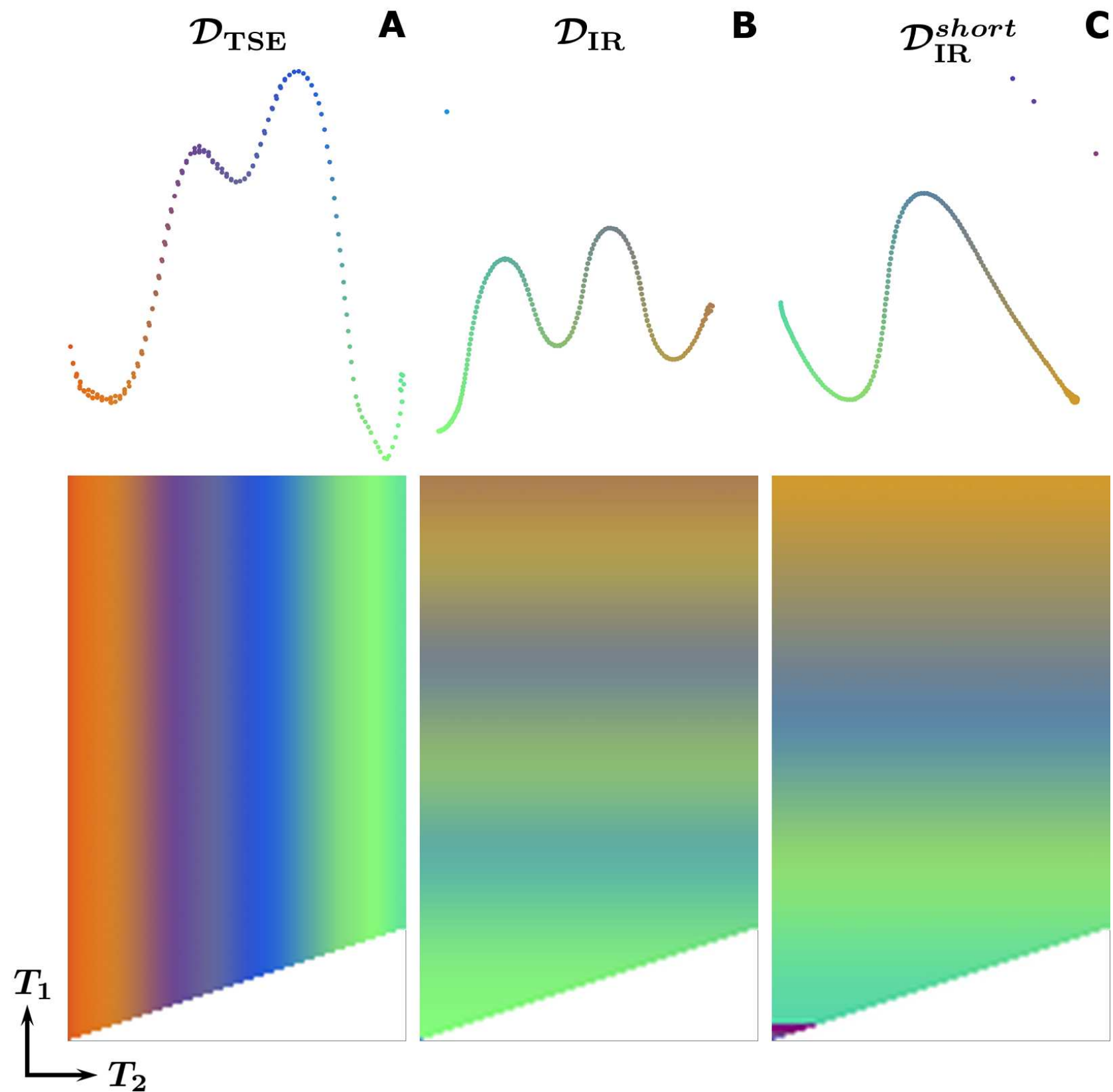}
\centering
\caption{\small Comparison of three classical sequences. Two-dimensional embeddings of the classical dictionaries (top) and the corresponding color-coded dictionary maps in the ($T_1$,\,$T_2$) coordinate system (bottom). In these dictionary maps, similar colors for certain ($T_1$,\,$T_2$) combinations indicate similar structure of the corresponding low-dimensional dictionary elements. (\textbf{A}) The different echo times in the TSE sequence enable $T_2$ encoding, resulting in a color variation only in the $T_2$ direction in $\dTSE$. (\textbf{B}) The opposite is true for the IR sequence, where different inversion times (TI$_{{max}}=3$ s) enable $T_1$ encoding, resulting in a color variation only in the $T_1$ direction in $\dIR$. (\textbf{C}) A shorter maximal inversion time (TI$_{{max}}=200$ ms) results in a reduced $T_1$ encoding, shown by the somewhat smaller color range in the color-coded dictionary map of $\dIR^{short}$ in the $T_1$ dimension. The white triangle in the bottom of the color-coded dictionary maps represents the unsampled region for which $T_2$ is larger than $T_1$.}
\label{fig:classical}
\end{figure}

\begin{figure}[!tb]
\centering
\includegraphics[width=\columnwidth]{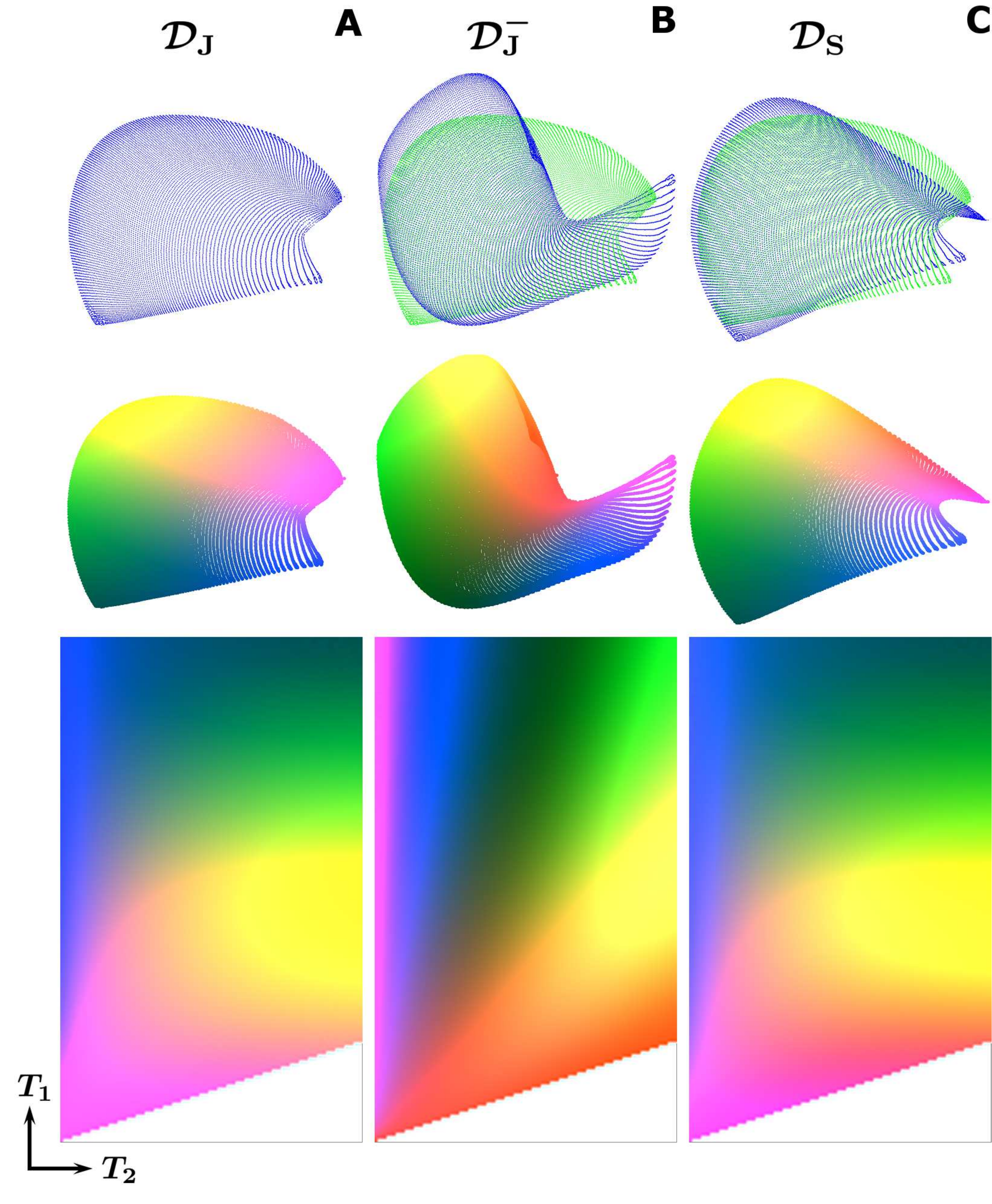}
\caption{\small Comparison of three different MRF flip angle patterns. Three-dimensional embeddings of $\dJ$ (\textbf{A}),  $\dJ^-$ (\textbf{B}) and  $\dS$ (\textbf{C}) (in blue), registered to that of $\dJ$ (in green) are shown in the top row, with their colored versions in the middle row. Embeddings for $\dJ$ and for $\dS$ look very similar, while both being very different from that for $\dJ^-$. The corresponding color-coded dictionary maps in the ($T_1$,\,$T_2$) coordinate system are shown in the bottom row, in which similar colors for certain ($T_1,T_2$) combinations indicate similar structure of the corresponding low-dimensional dictionary entries. Like the embeddings, also the color-coded dictionary maps for $\dJ$ and $\dS$ look very similar, suggesting comparable encoding capability. The sequence without inversion pulse results in a color-coded dictionary map with less color variation in the $T_1$ direction, especially for large $T_1$ values, suggesting reduced encoding capability compared to $\dJ$ and $\dS$. The white triangle in the bottom of the color-coded dictionary maps represents the unsampled region for which $T_2$ is larger than $T_1$.}
\label{fig:mrf}
\end{figure}

\begin{figure*}[!tb]
\includegraphics[width=.7\textwidth]{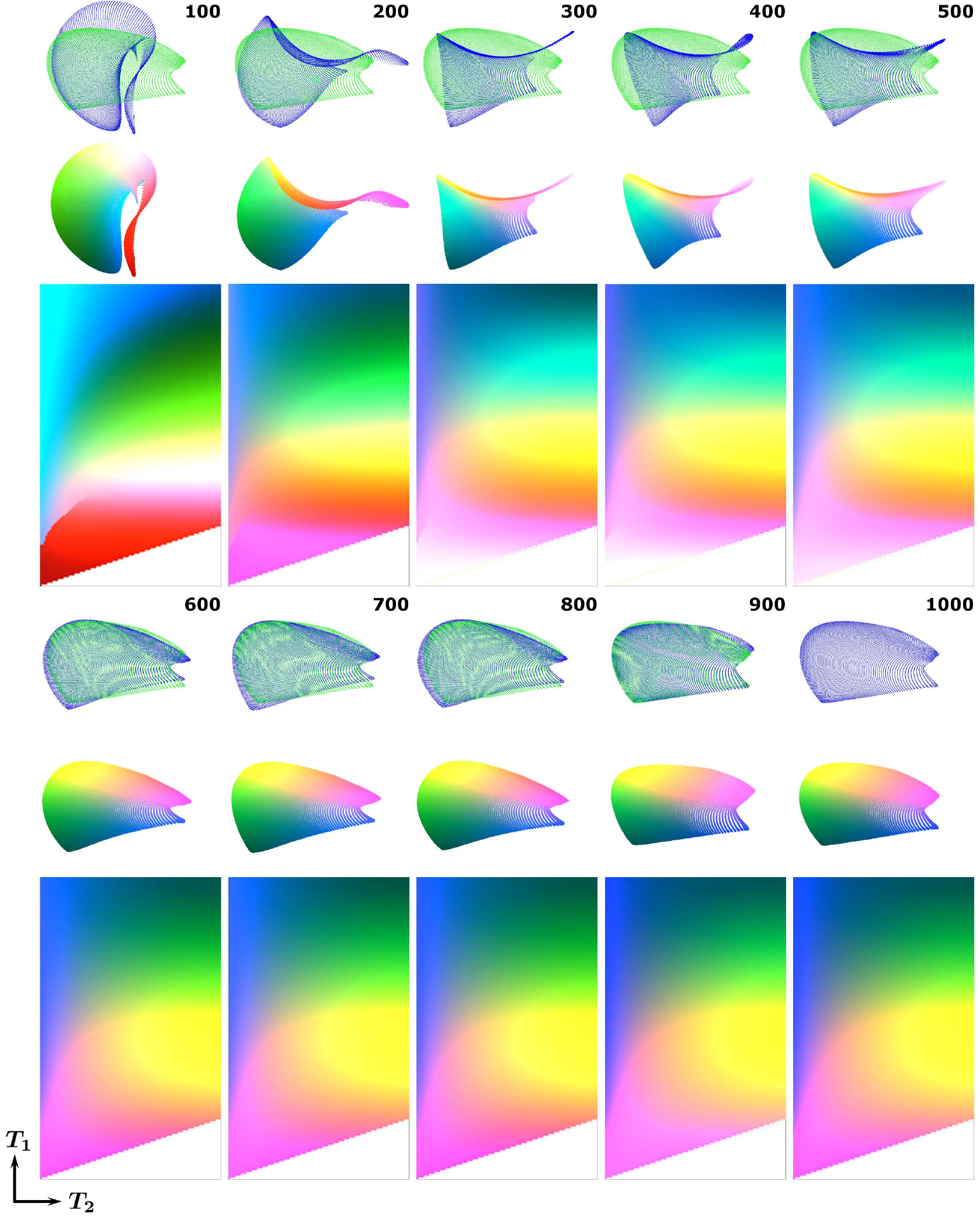}
\centering
\caption{\small Comparison of different MRF sequence lengths. Three-dimensional embeddings of $\dJ$ (top), their colored versions (middle), and the corresponding color-coded dictionary maps in the ($T_1$,\,$T_2$) coordinate system (bottom) for sequence lengths ranging from 100 to 1000 flip angles (lengths given on top of each subplot). For each different length the embedding (blue) was registered to that of length 1000 (green). There is less color variation in $T_2$ direction for the shortest ($L\in\{100,200\}$) compared to longer sequences, suggesting reduced $T_2$ encoding. From 600 flip angles onwards the embeddings and the color-coded dictionary maps are very similar to that of length 1000, in which case a clear color gradient can be observed both for $T_1$ and for $T_2$. The white triangle in the bottom of the color-coded dictionary maps represents the unsampled region for which $T_2$ is larger than $T_1$.}
\label{fig:lengths}
\end{figure*}

One of the parameters to tune in HSNE is the neighborhood size for the kNN search \cite{bib:Pezzotti2016}, which is related to the perplexity parameter of t-SNE \citep{bib:Maaten2013}. This parameter influences formation of clusters in the embedding and is dependent on the size of the data set (dictionary entries). Therefore, its value was empirically selected per case resulting in the following values: $3\cdot10^3$ (classical sequences), $10^3$ (MRF), and $5\cdot10^3$ ($B_1^+$ dictionary). For the classical sequences, such relatively high values of this parameter are motivated by the very low dependence of these sequences on either $T_1$ (TSE) or $T_2$ (IR). Hence, capturing more global structure of these dictionaries requires setting high values of this parameter. The neighborhood size for the other two cases were set to values of a similar order.

\subsection{Registration of embeddings}
To facilitate comparison between different embeddings, they were mapped to a common reference frame, ensuring consistency of the color mapping. Without loss of generality, we selected the embedding $\eJ$ corresponding to the dictionary $\dJ$ as the reference for the MRF sequences. For the comparison between different $B_1^+$ scaling factors each $\eJ^i$ for ${i\in \{0.4:0.1:1.3\}}$ was registered to $\eJ$, which coincides with embedding $\eJ^{1.0}$ corresponding to the subdictionary created with $B_1^+=1.0$. For the classical sequences introduced in Section~\ref{ssec:classical}, we registered $\eIR^{short}$ to $\eIR$. The registration was performed using a modification of the Iterative Closest Point (ICP) algorithm \citep{bib:Chen1992} by \citet{bib:Zinsser2005} that also enables scale estimation. In this we assumed that the correspondences between the point pairs are known, which allowed skipping the point matching step and significantly simplified the algorithm.

In our prior work \citep{bib:Dzyubachyk2019} we demonstrated, using Jiang's dictionary \citep{bib:Jiang2015} as an example, that embeddings produced using our approach exhibit a high degree of stability. This means that intrinsic stochastic effects resulting from using t-SNE/HSNE can be neglected. In the same work we also analyzed two ways of comparing two dictionaries: embedding them separately and jointly, in both cases followed by registration. Numerical results confirmed very similar performance of the two approaches, from which the conclusion was drawn that the former (separate embedding) is preferred for being faster. In this work we used separate embedding, followed by registration, for the MRF sequences, while two IR sequences were embedded together for better assessment of small differences in encoding ability.

\subsection{Color-coding of embeddings}
For each dictionary, we mapped the coordinates of the low-dimensional embedding into a color space. For 3D embeddings, the CIE L*a*b* color space \citep{bib:fairchild2013} was used. For 2D embeddings we used a colormap designed by \cite{bib:Teuling2011} that produces superior performance with respect to various validation measures compared to other 2D colormaps \citep{bib:Bernard2015}.
Consequently, the color of each entry was mapped back to the dictionary space. In this way a correspondence between each dictionary entry and a color was established, resulting in color-coded dictionary maps for each ($T_1$,\,$T_2$) or ($T_1$,\,$T_2$,\,$B_1^+$) combination. Typical examples of the color-coded embeddings and dictionary maps are illustrated in Figures~\ref{fig:classical}--\ref{fig:transmit-sep}. In these maps, similar colors indicate similar structure of the corresponding low-dimensional dictionary elements.

\begin{figure*}[!th]
\centering
\includegraphics[width=0.9\textwidth]{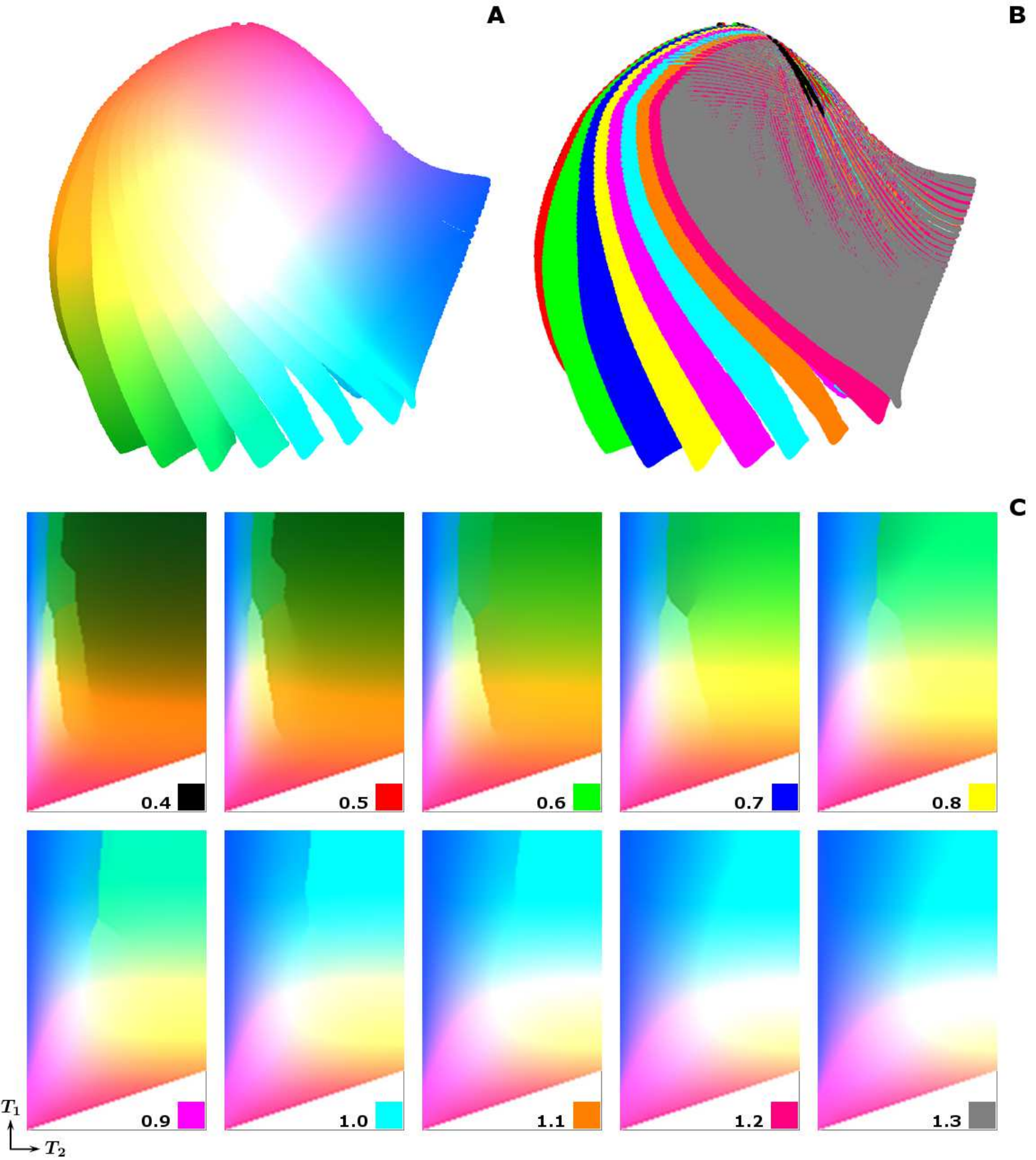}
\caption{\small Comparison of different $B_1^+$ scaling factors embedded together. (\textbf{A,B}) Three-dimensional embedding of $\dJ^B$ including $T_1$, $T_2$ and $B_1^+$ scaling factors ranging from 0.4 to 1.3. In (\textbf{A}), each embedding point is visualized using the developed color-coding scheme, while in (\textbf{B}) a single unique color was assigned to all points with the same $B_1^+$ value. From these plots it can be observed that different $B_1^+$ scaling factors tend to end up as separate subparts of the embedding, although the subparts have a common root. (\textbf{C}) The corresponding color-coded dictionary maps in the ($T_1$,\,$T_2$) coordinate system show that, in general, different $B_1^+$ fractions in the dictionary are represented by different colors in the color-coded dictionary maps, suggesting that the $B_1^+$ map can be estimated in the matching process together with the $T_1$ and $T_2$ maps. Some regions, such as represented by short $T_1$ and/or $T_2$, show very similar colors for different $B_1^+$ fractions, suggesting that these ($T_1$,\,$T_2$,\,$B_1^+$) combinations are less well-encoded by this particular MRF sequence. Please note that the colored squares in the white triangular regions refer to the different colors in (\textbf{B}). The white triangle in the bottom of the color-coded dictionary maps represents the region for which $T_2$ is larger than $T_1$. It is worth noting that sharp transitions visible in some of the maps are caused by much more complex structure of the embedding due to increased dictionary size.}
\label{fig:transmit}
\end{figure*}

\begin{figure*}[!th]
\centering
\includegraphics[width=0.8\textwidth]{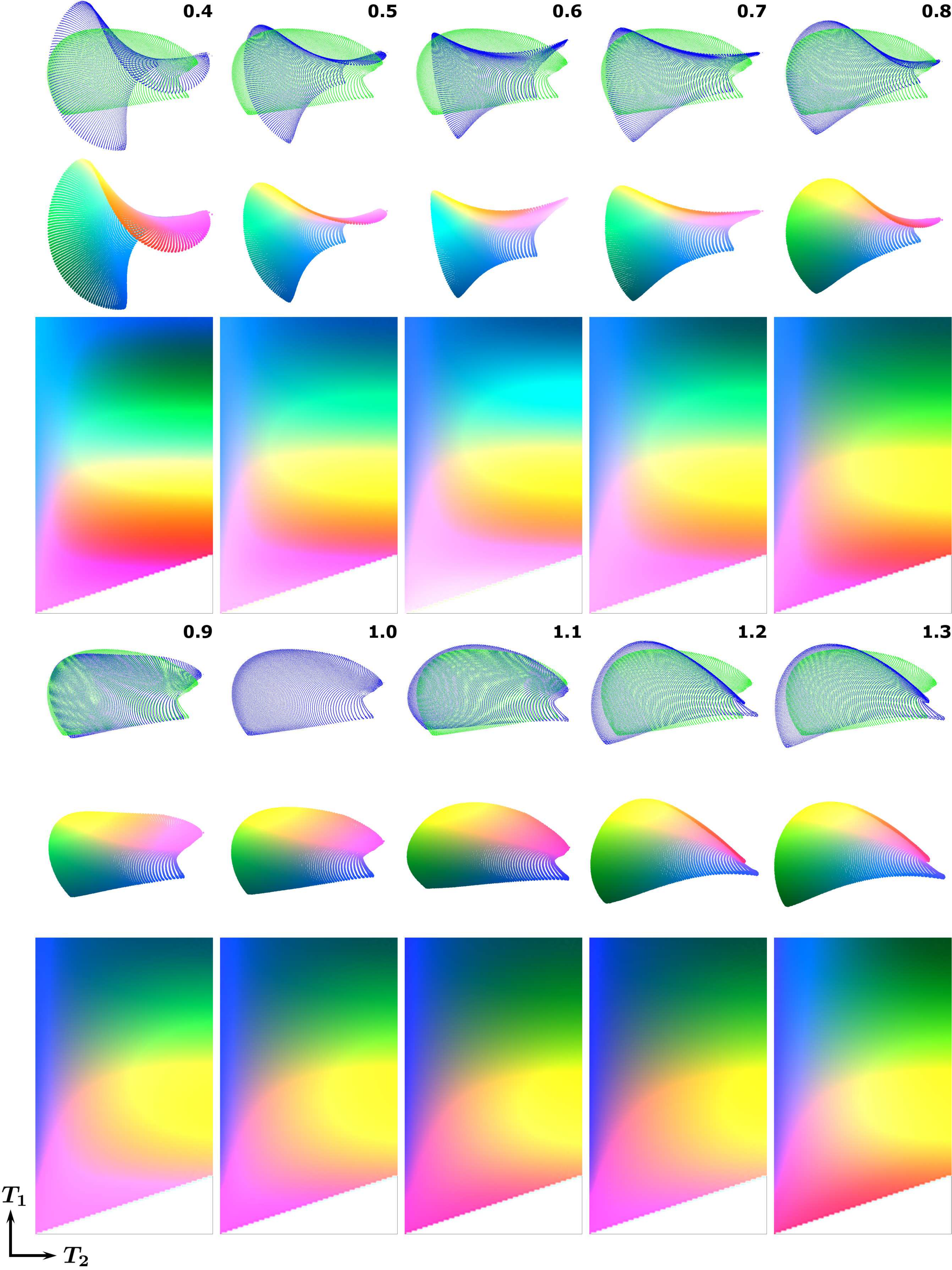}
\caption{\small Comparison of different $B_1^+$ scaling factors embedded separately. Three-dimensional embedding of $\dJ^B$ for $B_1^+$ scaling factors ranging from 0.4 to 1.3 (top row) and their colored versions (middle row). Subdictionaries were treated as individual dictionaries, embedded separately (blue) and registered to the embedding corresponding to $B_1^+=1.0$ (green) afterwards. From a $B_1^+$ scaling factor of approximately 0.9 onwards, the embeddings are very similar to that of $B_1^+$=1.0. The corresponding color-coded dictionary maps in the ($T_1$,\,$T_2$) coordinate system are shown in the bottom row with the $B_1^+$ fraction given above each subplot. For low $B_1^+$ fractions the color-coded dictionary maps shows less color variation in $T_2$ dimension compared to the $B_1^+=1.0$ case, especially for long $T_2$, suggesting reduced $T_2$ encoding for low $B_1^+$ fractions. The white triangle in the bottom of the color-coded dictionary maps represents the unsampled region for which $T_2$ is larger than $T_1$.}
\label{fig:transmit-sep}
\end{figure*}

\subsection{Simulation experiment}
A synthetic image of the brain \citep{bib:Guerquin-Kern2012} was resized to a $256\times256$ matrix and binned into three tissue types representing white matter (WM), gray matter (GM) and cerebral spinal fluid (CSF). For each of these three tissues, relaxation time values reported in literature \citep{bib:Wansapura1999} were assigned to the corresponding regions, resulting in ground truth $T_1^{\rm{true}}$ and $T_2^{\rm{true}}$ maps. MRF and IR signals were simulated for the synthetic brain image by assigning to each pixel the dictionary entry corresponding to the ($T_1^{\rm{true}}$,\,$T_2^{\rm{true}}$) or ($T_1^{\rm{true}}$,\,$T_2^{\rm{true}}$,\,$B_1^+$$^{\rm{true}}$) combination. Noise was added to these signals such that the resulting SNR of an MRF signal curve was equal to 1. The simulated MRF and IR signals were then matched back to the dictionary according to
\begin{equation}
{m} = \argmax_{i} \{ \mathbf{d}_i \cdot \mathbf{s} \}
\end{equation}
with $\mathbf{d}_i=\mathbf{d}_i(x,y)$ denoting the normalized dictionary entries, $\mathbf{s}=\mathbf{s}(x,y)$ the normalized MRF signal, and ${m}=m(x,y)$ the index pointing to the best match for each pixel $(x,y)$. Note that this approach results in a $T_1$ map for the IR sequences, $T_1$ and $T_2$ maps for MRF sequences, and $T_1$, $T_2$ and $B_1^+$ maps for $\dJ^{B}$. In addition, the matching was repeated for $\dJ^{B}$, this time treating the $B_1^+$ value as a known fixed spatially-invariant value, which is relevant when a $B_1^+$ map is measured beforehand. Note that the same experiment can be performed for a spatially-variant $B_1^+$ map. Error maps for $T_1$ and $T_2$ were calculated from
\begin{equation} \label{E1}
E_{(1:2)}(x,y) = \frac{\left|{T}_{(1:2)}(x,y)-{T}_{(1:2)}^{\rm{true}}(x,y)\right|}{{T}_{(1:2)}^{\rm{true}}(x,y)}\cdot 100\%.
\end{equation}
Matching errors for gray matter, white matter and CSF were derived from
\begin{equation}\label{Eav1}
\overline{E}_{(1:2)}= \frac{\left|\overline{{T}}_{(1:2)}-\overline{{T}}_{(1:2)}^{\text{true}}\right|}{\overline{{T}}_{(1:2)}^{\text{true}}}\cdot 100\%,
\end{equation}
where $\overline{{T}_1}$ and $\overline{{T}_2}$ are the $T_1$ and $T_2$ maps averaged over all pixels in the corresponding regions.

\section{Results}

\subsection{Classical sequences}
\label{ssec:results-classical}
Figure~\ref{fig:classical} shows the low-dimensional (2D) representations and the corresponding color-coded dictionary maps for the dictionaries generated for the three classical sequences: $\dTSE$, $\dIR$ and $\dIR^{short}$. These classical sequences only encode one parameter each, and therefore the embeddings have relatively simple structure (one-dimensional manifolds). The multi-echo TSE sequence used for $T_2$ mapping only encodes $T_2$, which is confirmed in Figure~\ref{fig:classical}A by the color change in the $T_2$ direction, while there is no color change in the $T_1$ direction. The opposite is true for the $T_1$ mapping IR sequence illustrated in  Figure~\ref{fig:classical}B. Shortening the maximal inversion time in this sequence reduces the $T_1$ encoding ability, as represented by the smaller color-range in Figure~\ref{fig:classical}C compared to that in Figure~\ref{fig:classical}B. These results are confirmed by the simulation results in Figure~\ref{fig:classical_sim}. 

\begin{figure*}[!tb]
\centering
\hbox{\includegraphics[width=\columnwidth]{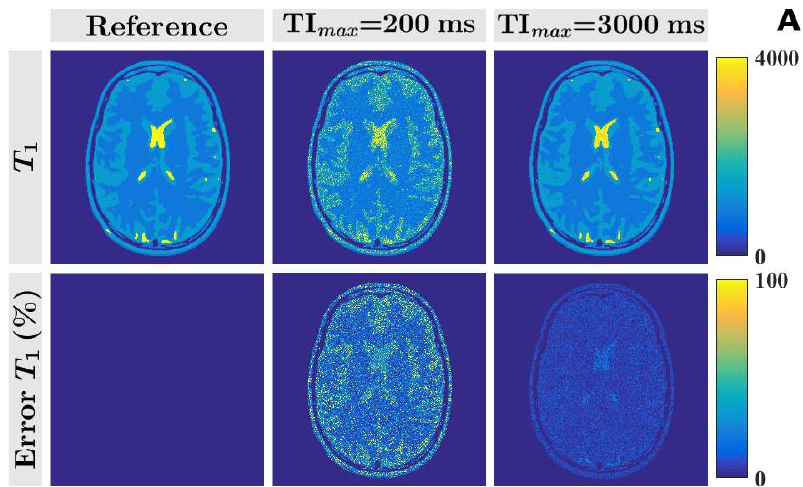}\hspace{0.5cm}\hfill
\includegraphics[width=\columnwidth]{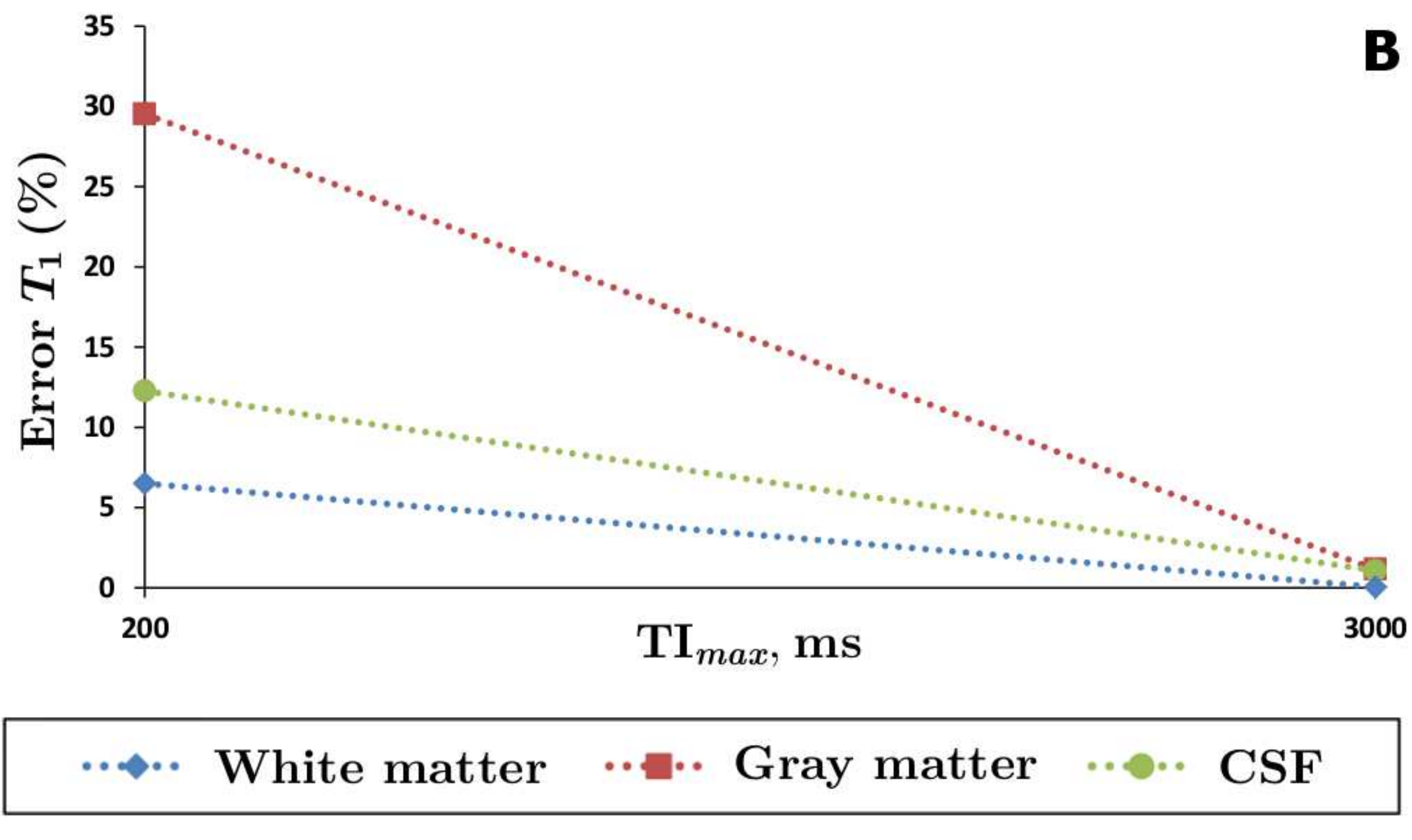}}
\caption{\small Simulation study of the encoding capability for the classical IR sequence with different maximal inversion times TI$_{{max}}$. (\textbf{A}) The $T_1$ map obtained from $\dIR$ is of higher quality compared to that of $\dIR^{short}$, confirmed by the percentage error maps calculated according to Equation~\eqref{E1}. (\textbf{B}) Percentage errors in $T_1$ averaged over the entire individual tissue components according to Equation~\eqref{Eav1} show that the encoding difference is larger for gray matter than for white matter and CSF. }
\label{fig:classical_sim}
\end{figure*}

\subsection{MRF sequences}
\label{ssec:results-MRF}
Figure~\ref{fig:mrf} shows the embeddings and the color-coded dictionary maps for the dictionaries generated for the three MRF sequences: $\dJ$, $\dJ^-$ and $\dS$. These sequences encode $T_1$ and $T_2$ simultaneously, resulting in embeddings that are more complicated structures (three-dimensional manifolds) compared to those of the classical sequences. The differences between the three MRF sequences are much smaller compared to the differences between the classical sequences. Removing the inversion pulse from Jiang's sequence reduces the encoding capability, which can be observed from the smaller color variation in the $T_1$ direction in the color-coded dictionary map, especially for long $T_1$ values. The embeddings and the color-coded dictionary maps for Sommer's and Jiang's sequence look very similar, suggesting that those sequences provide comparable encoding quality. These results are confirmed by the simulation results presented in Figure~\ref{fig:mrf_sim}, where the largest matching error is found for $\dJ$, especially for tissues with very long $T_1$ values such as CSF.

\begin{figure*}[!tb]
\centering
\includegraphics[width=\columnwidth]{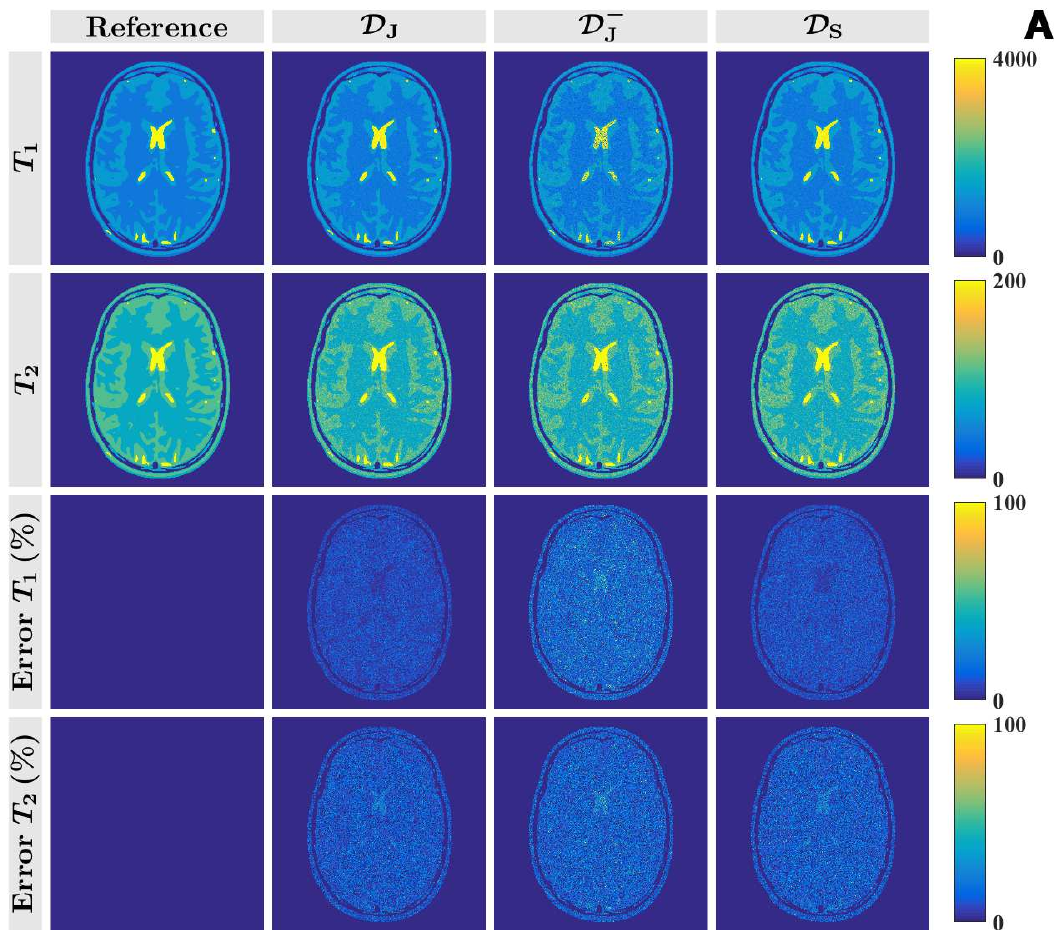}\hfill
\includegraphics[width=\columnwidth]{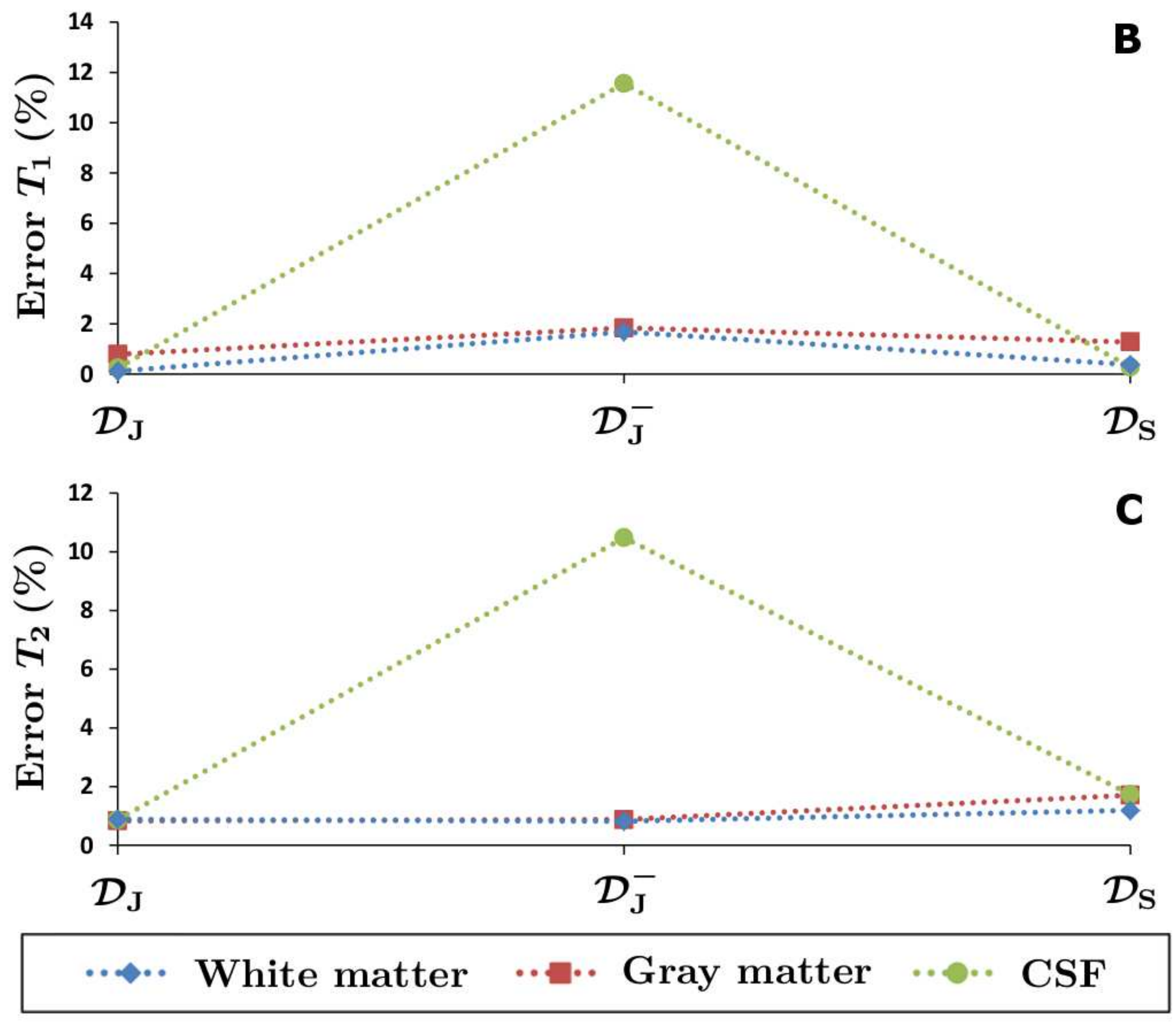}
\caption{\small Simulation study of the encoding capability for the three studied flip angle patterns. (\textbf{A}) The $T_1$ and $T_2$ maps obtained from $\dJ$ are of very similar quality to that of $\dS$, confirmed by the percentage error maps calculated according to Equation~\eqref{E1}. The absence of the inversion pulse in $\dJ^-$ reduces the encoding capability compared to $\dJ$. (\textbf{B,C}) Percentage errors in $T_1$ and $T_2$ averaged over the entire individual tissue components according to Equation~\eqref{Eav1} show that the encoding difference is largest for tissues with very long $T_1$ values such as is the case for CSF. The lines connecting the markers are shown for visualization purpose only and have no other meaning.}
\label{fig:mrf_sim}
\end{figure*}

\subsection{Sequence length}
\label{ssec:results-length}
Figure~\ref{fig:lengths} shows the 3D embeddings and the color-coded dictionary maps for $\dJ$ and its truncated versions, corresponding to different flip angle sequence lengths. For a very small number of flip angles, $T_2$ encoding is reduced, shown by less color variation in the $T_2$ direction. At 600 and above flip angles, the embeddings and the color-coded dictionary maps start looking very similar to that of the full-length version. These results are also confirmed by simulation results presented in Figure~\ref{fig:lengths_sim}, which shows a larger matching error for shorter sequence lengths as was also shown by \citet{bib:Sommer2017}. This effect is especially visible for $T_2$, for which the encoding principle relies on stimulated echoes whose contribution becomes larger for longer sequences. The $T_2$ matching error is smallest for tissues with long $T_1$ and $T_2$ (CSF), which is also predicted by the larger color gradient in the $T_2$ direction in the color-coded dictionary maps for long $T_1$ values, which is especially visible for a sequence length of 100.

\begin{figure*}[!tb]
\centering
\includegraphics[width=\textwidth]{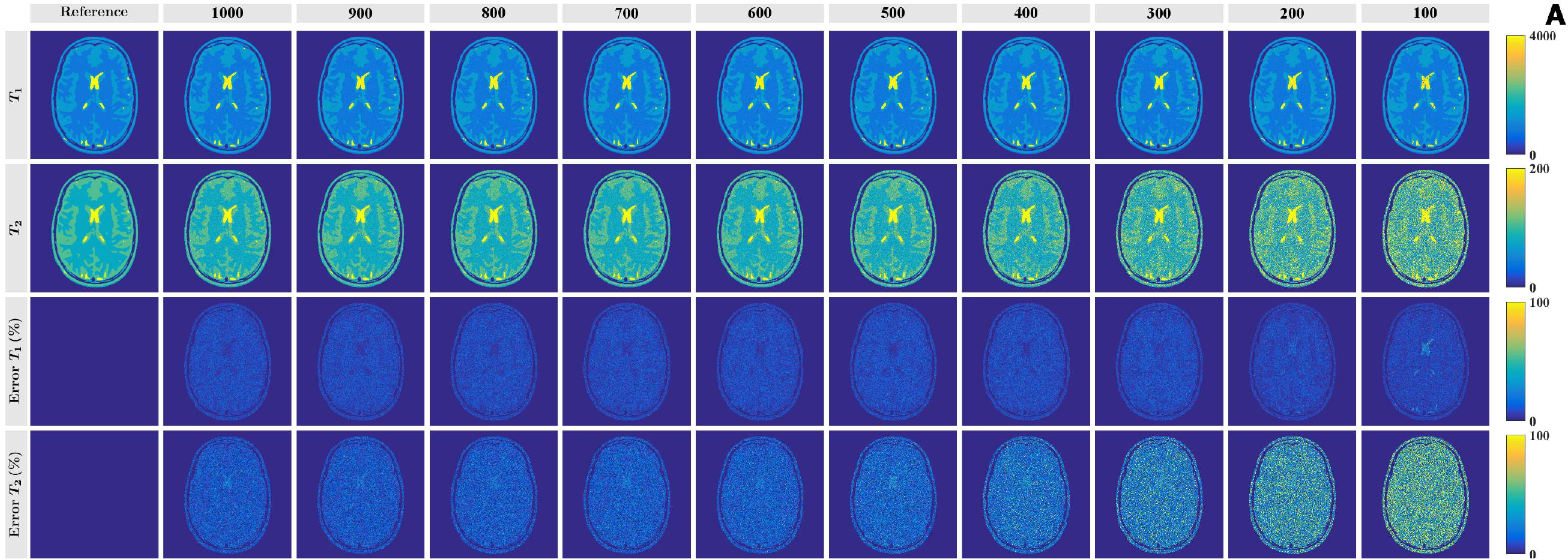}

\vspace{0.5cm}

\includegraphics[width=\textwidth]{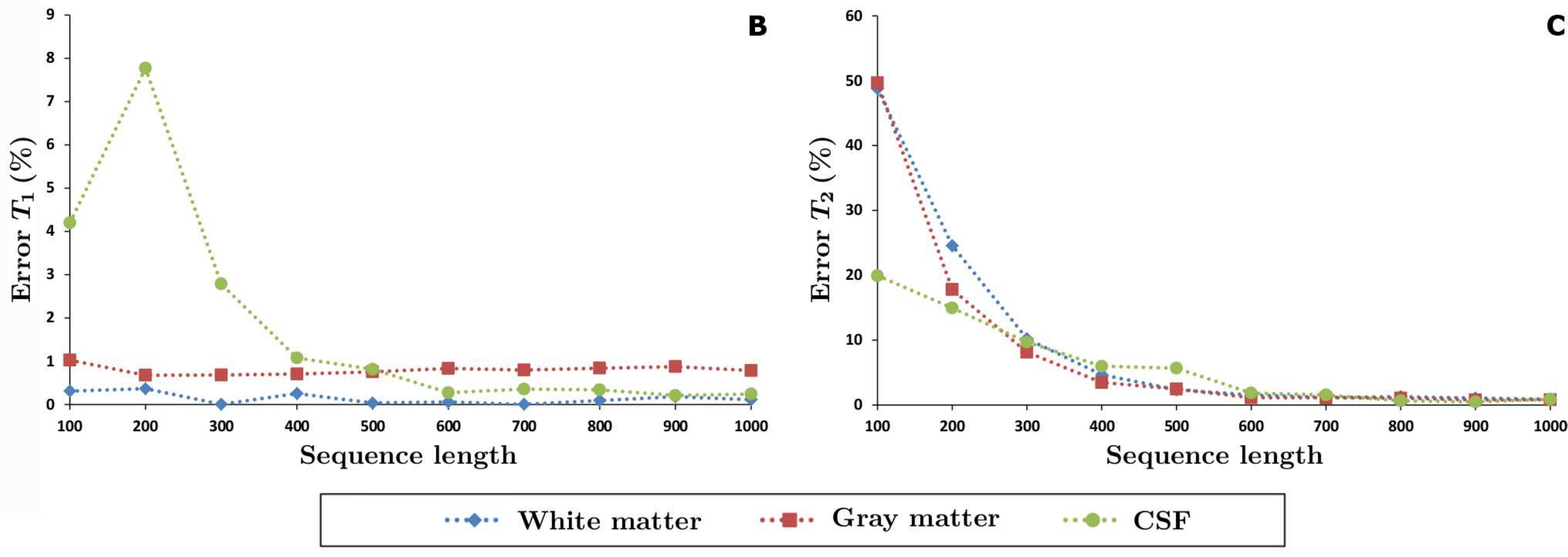}
\caption{\small Simulation study of the encoding capability for different sequence lengths. (\textbf{A}) The $T_2$ maps obtained from $\dJ$ show only a minor increase in error (Equation~\eqref{E1}) as the length of the dictionary elements reduces from 1000 to 700 flip angles. Further shortening the length of the dictionary elements to 100 drastically increases the error in $T_2$. This effect is hardly visible for $T_1$, again, except for tissues with very long $T_1$ values. (\textbf{B,C}) These results are also summarized in the average percentage errors calculated over the entire individual tissue components according to Equation~\eqref{Eav1}. }
\label{fig:lengths_sim}
\end{figure*}

\subsection{$B_1^+$ scaling factors}
\label{ssec:results-B1}
Figure~\ref{fig:transmit} shows the embedding and the color-coded dictionary map for $\dJ^{B}$, in which different $B_1^+$ scaling factors represent transmit field inhomogeneities or inefficiencies. Note that the entire dictionary including $B_1^+$ variations was embedded as one single dictionary, from which $B_1^+$ can be estimated together with $T_1$ and $T_2$ in the matching process. The color-coded dictionary maps show a gradual color change for different $B_1^+$ scaling factors. This color gradient appears to be smaller for short ($T_1$,\,$T_2$) combinations, suggesting lower encoding for those regions. The color gradient appears also smaller in the $T_2$ direction, suggesting larger matching errors for $T_2$ than for $T_1$. Simulation results in Figure~\ref{fig:transmit_sim} confirm these findings, showing a larger matching error for $T_2$ than for $T_1$, and the error is also larger for gray matter and white matter that have shorter $T_1$ and $T_2$ values compared to CSF.

\begin{figure*}[!tb]
\centering
\includegraphics[width=\textwidth]{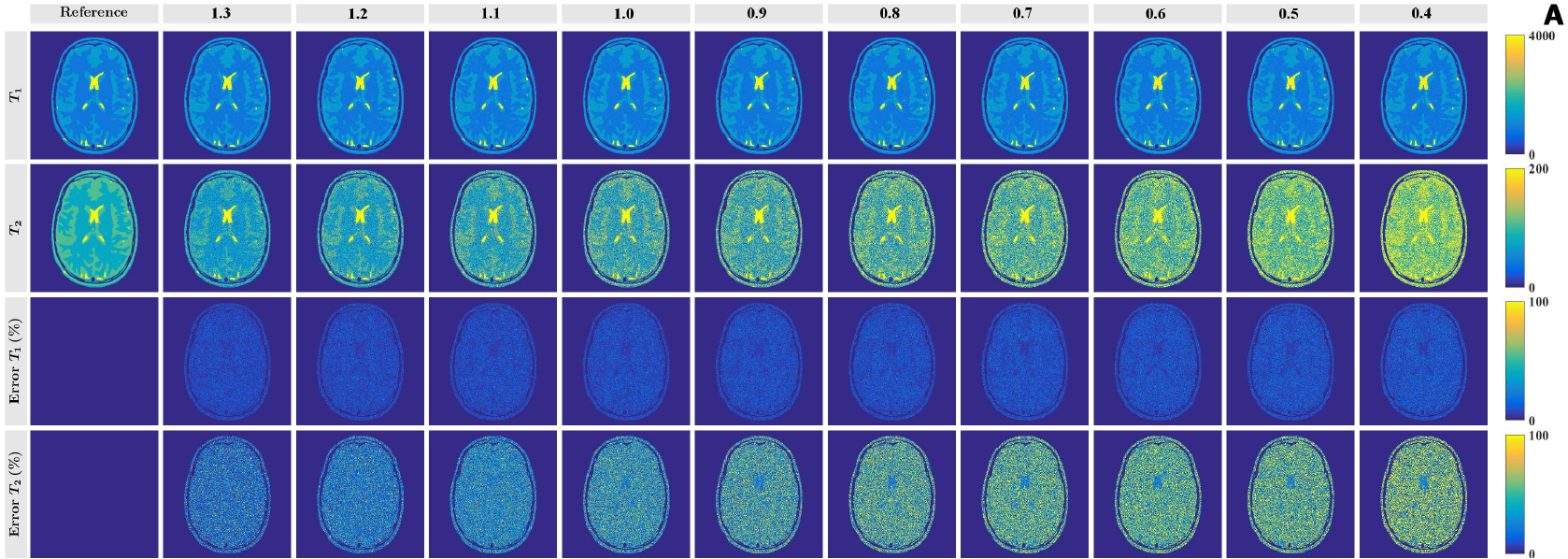}

\vspace{0.5cm}

\includegraphics[width=\textwidth]{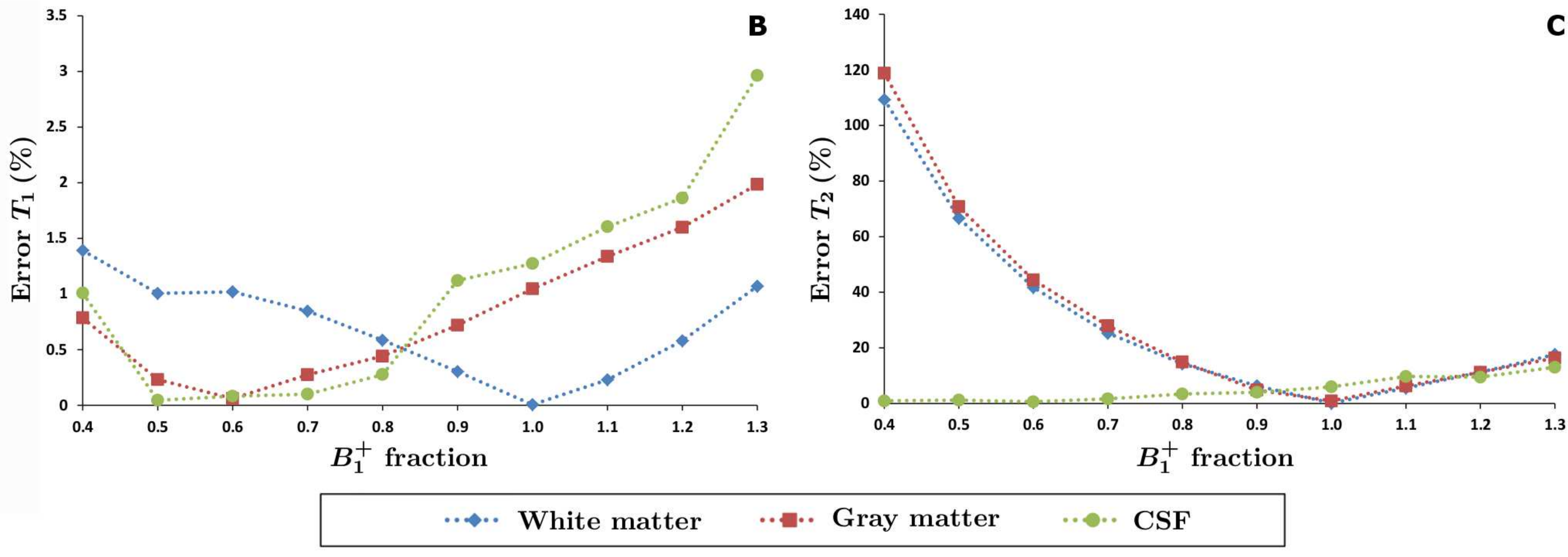}
\caption{\small Simulation study of the encoding capability for $\dJ^{B}$ when matching $T_1$, $T_2$ and $B_1^+$. (\textbf{A}) The error in the $T_2$ maps increases as the $B_1^+$ scaling factor, and therefore the maximum flip angle, decreases. (\textbf{B,C}) This effect is shown again in the percentage errors averaged over the entire individual tissue components (Equation~\eqref{Eav1}), which also show that the effect is very small for $T_1$. A scaling factor of $B_1^+=1.0$ results in the smallest error for white matter and gray matter.}
\label{fig:transmit_sim}
\end{figure*}

Figure~\ref{fig:transmit-sep} shows the embeddings and color-coded dictionary maps for subdictionaries of $\dJ^{B}$, each embedded individually. For small $B_1^+$ scaling factors the color-coded dictionary map shows less color variation in the $T_2$ dimension, suggesting reduced $T_2$ encoding. This effect is strongest in the right half of the color-coded dictionary maps, corresponding to long $T_2$ values. For $B_1^+$ scaling factors between 0.8 and 1.3 the embeddings and the color-coded dictionary maps look rather similar, predicting smaller matching errors for high $B_1^+$ values than for low values. These results are confirmed by simulation results in Figure~\ref{fig:transmit_sim-sep}, where the percentage errors are larger for $T_2$ than for $T_1$. They are furthermore especially pronounced for tissues with long $T_2$ values (CSF) and for low $B_1^+$ scaling factors. In general these matching errors are much larger than those in Figure~\ref{fig:transmit_sim}, showing the advantage of fixing the $B_1^+$ value in the matching process.

\begin{figure*}[!tb]
\centering
\includegraphics[width=\textwidth]{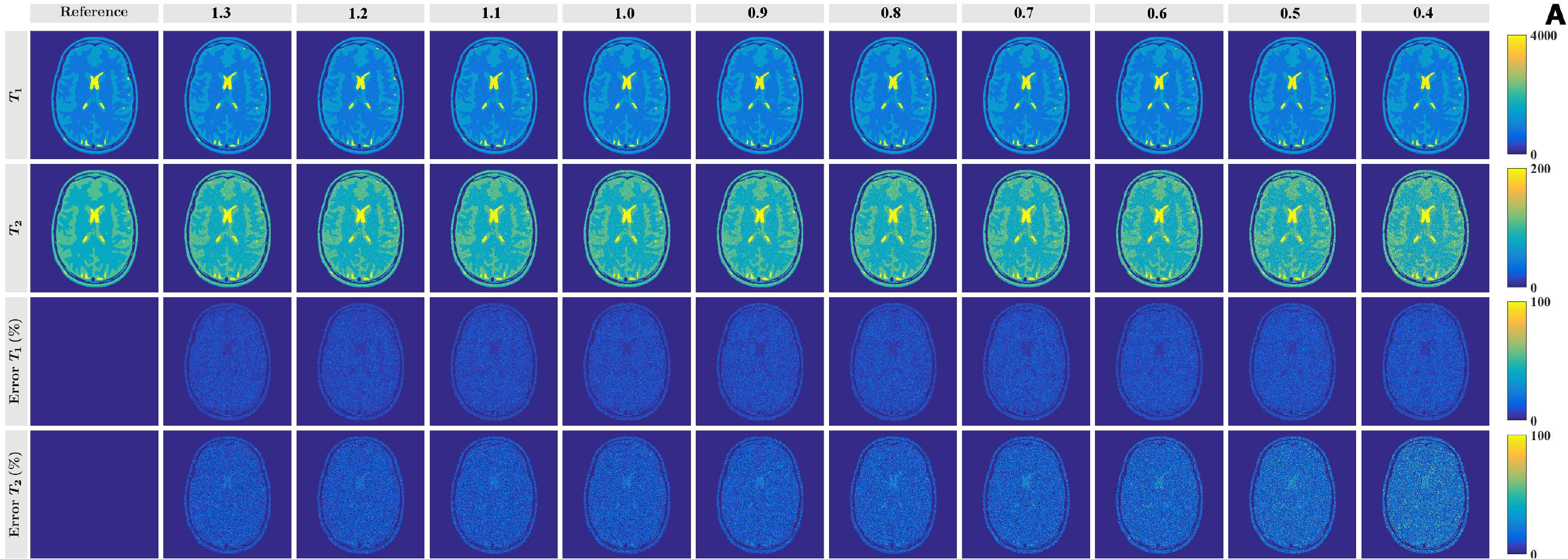}

\vspace{0.5cm}

\includegraphics[width=\textwidth]{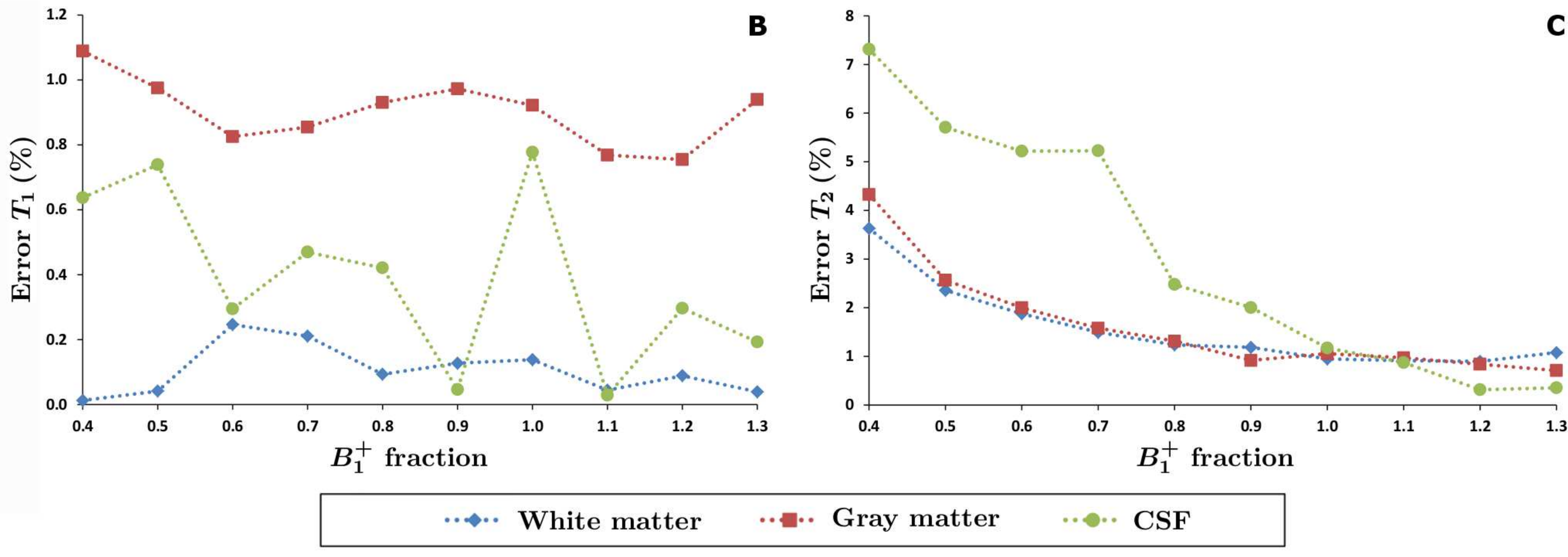}
\caption{\small Simulation study of the encoding capability for $\dJ^{B}$ when matching $T_1$ and $T_2$, but not $B_1^+$. (\textbf{A}) The error in the $T_2$ maps increases as the $B_1^+$ scaling factor, and therefore the maximum flip angle, decreases. The error in the $T_1$ map is not sensitive to $B_1^+$ scaling variations. (\textbf{B,C}) This effect is shown again in the percentage errors averaged over the entire individual tissue components (Equation~\eqref{Eav1}), which also show that the effect is very small for $T_1$. The matching error is largest for tissues with long $T_2$ values such as CSF.}
\label{fig:transmit_sim-sep}
\end{figure*}

\section{Discussion}

This work has shown the feasibility of using HSNE to visualize and compare the encoding capability of different quantitative sequences. Low-dimensional representations of the classical sequences showed color variation either in the $T_1$ or $T_2$ direction, whereas MRF sequences showed color variations in both the $T_1$ and $T_2$ directions. A very short MRF sequence results in reduced $T_2$ encoding, and from a length of 600 flip angles onwards the encoding capability of Jiang's pattern \citep{bib:Jiang2015} is very similar to that of a sequence with a length of 1000 flip angles. Different $B_1^+$ values are, in general, well distinguishable, except for very small $T_1$ and/or $T_2$ combinations. When using the $B_1^+$ maps as prior information in the matching process, the encoding is better for short $T_2$ tissues than for long $T_2$ tissues. All these results were in agreement with IR and MRF simulation results, showing additionally that fixing the $B_1^+$ in the matching process results in smaller errors compared to estimating $B_1^+$ in the matching process.

Although in this study t-SNE/HSNE was used to transform the high-dimensional dictionaries into low-dimensional space, there are many other dimensionality reduction techniques that could be used instead. For several benchmark data sets, t-SNE has been shown to produce results of superior quality compared to other non-linear transformations such as Isomap and Locally Linear Embedding \citep{bib:vanDerMaaten2008}.  However, these results need to be reevaluated for MR dictionaries in order to find out whether these conclusions hold in the context of quantitative MR sequences. 

The results shown in this study provided qualitative comparisons of different MR sequences by visualizing differences using color-coding. Although these qualitative results were in agreement with simulation results and therefore show the proof of principle of the technique, quantitative results derived from the color-coded dictionary maps or the low-dimensional embeddings may facilitate better and more detailed comparisons between sequences that show small differences in encoding capability. One could describe the local shape of the embeddings using e.g. statistics of the point cloud distribution. Currently the question of which quantitative measures are suitable for describing such differences most efficiently, or which measures describe the encoding capability best, is still open. 

In our earlier work \citep{bib:Dzyubachyk2019} we presented two different ways to perform the embedding and registration processes when comparing different sequences. In that work we embedded each sequence individually, after which the point clouds were registered to each other using a modification of the Iterative Closest Point algorithm with integrated scale estimation \citep{bib:Zinsser2005}. An alternative approach would be to first combine two dictionaries and treat them as one large dictionary in the embedding process, after which registration of the two point clouds corresponding to each of the combined dictionaries can be performed. As we demonstrated earlier in \citet{bib:Dzyubachyk2019}, both approaches provide comparable registration results, and hence result in similar color-coded dictionary maps. Since the latter approach is computationally more expensive due to the two-fold increase in the number of high-dimensional data points, the former approach is more attractive in this application.

The HSNE algorithm is intrinsically stochastic due to random initialization of the low-dimensional distribution. Hence, the final embeddings may in general differ from run-to-run. In \citet{bib:Dzyubachyk2019} we performed a quantitative stability study on the Jiang's dictionary \citep{bib:Jiang2015} and concluded that the final embedding was highly repeatable. While we did not repeat this stability study for different dictionaries used in this paper, we performed several runs on the most characteristic dictionaries from each group and qualitatively concluded sufficient similarity between the results corresponding to different runs of the algorithm. The perplexity, the only variable HSNE parameter in our setup, depends on the number of points in the data set and  was thus optimized per dictionary group. It must also be pointed out that the $T_1$ and $T_2$ range included in the dictionary also affects the final embedding structure. Therefore, HSNE parameter(s) should ideally be optimized for each dictionary with a different size or parameter range.


In visualization applications, special attention should de paid to selection of the color scheme \citep{bib:Bernard2015}. While this is rather straightforward in 3D, as most of the commonly used colormaps are perceptually linear, selection of a proper 2D colormap is much less trivial as such colormaps should comply to a large set of quality requirements as pointed out by \citet{bib:Bernard2015}. Based on their experiments they concluded that the 2D colormap designed by \citet{bib:Teuling2011} outperforms all other analyzed colormaps. Consequently, we selected this colormap for visualization for cases in which the dictionary was projected onto 2D.

High-dimensional dictionaries can be transformed into any $n$-dimensional space using t-SNE for $n$ smaller than the dynamic length of the dictionary. Natural choices enabling straightforward visualization are $n=2$ or $n=3$. In this work, all MRF dictionaries were transformed into a 3D space, since 3D embeddings are expected to show larger differences between dictionaries if the differences between encoding capability are relatively small \citep{bib:Koolstra2019}. However, the classical sequences described in Section~\ref{ssec:classical} exhibit very low dependence on either $T_1$ or $T_2$. This means that the corresponding embedding will always lay on a 1D manifold, irrespective of the target embedding dimensionality of the low-dimensional space. Thus, in this work we projected these sequences onto a 2D space, which was preferred to 1D embedding for ease of visualization and being more commonly used, as dimensionality reduction to 1D remains very rare in visualization applications and its properties are not well known or/and described.

\section{Conclusion}

HSNE can be used to visualize the encoding capability of classical quantitative sequences and MRF sequences. The technique can be used to obtain insight into the encoding principles, in particular of MRF, by comparing different sequences. The framework may furthermore be helpful for MRF sequence optimization, in which the application of interest and its corresponding constraints can easily be taken into account. Further work needs to be done to derive quantitative measures of encoding capability from low-dimensional embeddings, which may support the use of thic technique in clinical applications.   

\section*{Acknowledgments}
This project was funded by the European Research Council Advanced grant 670629 NOMA MRI, and partially by The Netherlands Technology Foundation (STW) as part of the STW project 12721 (Genes in Space) under the Imaging Genetics (IMAGENE) Perspective program.


\bibliographystyle{model2-names}
\bibliography{journals_abbrev,refs}

\end{document}